\documentclass[12pt]{iopart}
\usepackage{iopams}
\usepackage{graphicx}

\begin{document}

\title[Phonon number quantum jumps in an optomechanical system]{Phonon number quantum jumps in an optomechanical system}

\author{A A Gangat, T M Stace and G J Milburn}
\address{Department of Physics, School of Mathematics and Physics, The University of Queensland, St Lucia, QLD 4072, Australia}
\ead{a.gangat@physics.uq.edu.au}


\begin{abstract}
We describe an optomechanical system in which the mean phonon number of a single mechanical mode conditionally displaces the amplitude of the optical field.  Using homodyne detection of the output field we establish the conditions under which phonon number quantum jumps can be inferred from the measurement record: both the cavity damping rate and the measurement rate of the phonon number must be much greater than the thermalization rate of the mechanical mode.  We present simulations of the conditional dynamics of the measured system using the stochastic master equation.  In the good-measurement limit, the conditional evolution of the mean phonon number shows quantum jumps as phonons enter and exit the mechanical resonator via the bath.
\end{abstract}

\section{Introduction}
Observation of energy quantization in a macroscopic mechanical oscillator mode is of fundamental interest in exploring the quantum-to-classical transition.  Electromechanical systems, where the relevant mechanical mode interacts with devices such as superconducting quantum interference devices, single-electron transistors, cooper-pair boxes, or even other mechanical modes, provide a possible route toward achieving this, and various scenarios have been explored theoretically \cite{IrishSchwab}-\cite{Woolley}.  Experimental realization, however, remains a challenge.  A second platform where observation of macroscopic mechanical mode energy quantization may become possible is optomechanics \cite{optomechanics-review1, optomechanics-review2}.   In an optomechanical system, a collective vibrational mode of a mesoscopic system, such as a doubly clamped beam \cite{Groblacher}, a ring \cite{Vahala}, a suspended membrane ("membrane-in-the-middle") \cite{Thompson, Sankey}, or even a cloud of ultracold atoms \cite{Murch}-\cite{Purdy}, is parametrically coupled to an electromagnetic cavity field mode.  While also challenging, there has been much interest in the prospects of experimental observation of macroscopic mechanical mode energy quantization in an optomechanical system due to recent developments.  In particular, direct coupling of the optical cavity mode to the square of the mechanical displacement coordinate has been experimentally demonstrated in the "membrane-in-the-middle" system \cite{Thompson, Sankey} and its ultracold atoms analogue \cite{Purdy}.  Under the rotating wave approximation, such a coupling allows for a quantum nondemolition (QND) \cite{WM}  measurement of the mechanical mode phonon number.  Observation of mechanical energy quantization in such a system can then be achieved either indirectly, through phonon shot noise measurement, or directly, by detecting {\em quantum jumps} in the phonon number via continuous measurement of the cavity field.  A. A. Clerk {\em et al.} have analyzed the former case in \cite{Clerk}.  In this paper we present an analysis of the latter.  In particular, we show the nature of the displacement of the cavity field in the strongly driven regime in response to changes in the phonon number, and we establish constraints on the parameter space which must be satisfied in order for {\em quantum jumps} of the phonon number to be observed.

There are two ways we can describe a measurement in quantum mechanics. Firstly we can simply give the solution to the master equation describing the interaction between the measured system and the apparatus and any decay channels that may be present. This is called the {\em unconditional evolution}. Secondly, in the case of a continuous measurement we wish to describe the {\em conditional} state of the measured system given a particular classical stochastic record for the measurement. For the case considered here this will require a phase-sensitive measurement of the cavity field, such as homodyne detection. The measurement record is then a stochastic homodyne current and the evolution of the mechanical resonator is conditioned on a particular realization of this current \cite{WM}. Of course the statistics of the homodyne current itself is partly determined by the quantum state of the mechanical resonator.  The goal is then to reconstruct the quantum trajectory of the mechanical mode by suitably filtering the measurement record, and to observe phonon absorptions and emissions of the mechanical resonator mode as {\em retroactive quantum jumps} \cite{MabuchiWiseman} in the trajectory.  If one averages the conditional state of the system over all possible measurement records, one obtains the unconditional state. 

\section{The model}
We assume the case of a nearly ideal single-port cavity such that the constraint established by Miao {\em et al.} in \cite{Miao} is satisfied.  Under a single-mode description of the cavity the generic Hamiltonian for the optomechanical system with a drive on the cavity is
\begin{equation}
H=\hbar\omega_{c}(x_0+x_b)a^\dagger a+\hbar\nu b^\dagger b+\hbar(\epsilon^* a e^{i\omega t}+\epsilon a^\dagger e^{-i\omega t}),
\end{equation}
where $a,a^\dagger$ refer to the optical mode and $b,b^\dagger$ refer to the mechanical mode,  $\omega_c(x)$ is the cavity mode frequency dependent on the mechanical resonator position $x$, $\nu$ is the frequency of the mechanical mode, $\omega$ and $\epsilon$ refer respectively to the frequency and amplitude of the coherent drive field on the cavity, $x_0$ is the equilibrium position of the mechanical resonator, and
\begin{equation}
x_b=\sqrt{\frac{\hbar}{2\nu m}}(b+b^\dagger)
\end{equation}
is the displacement coordinate of the mechanical mode.  If the mechanical resonator is positioned such that $x_0$ is an extremum of $\omega_c(x)$, we approximate
\begin{equation}
H=\hbar\omega_{c}(x_0)a^\dagger a+\hbar\nu b^\dagger b+\hbar(\epsilon^* a e^{i\omega t}+\epsilon a^\dagger e^{-i\omega t})
+\frac{\hbar}{2} G a^\dagger a(b+b^\dagger)^2,
\label{ham2}
\end{equation}
where
\begin{equation}
G=\frac{\hbar}{2\nu m}\left . \frac{\partial^2\omega_c(x)}{\partial x^2}\right |_{x=x_0}.
\end{equation}

We now assume that the optical cavity remains near the steady state it would have reached if $G=0$. In an interaction picture at the driving frequency, this is a coherent state with amplitude $\alpha_0=-\frac{i\epsilon}{\kappa/2+i\delta}$, where $\delta=\omega_c-\omega$ is the detuning between the cavity driving field and the cavity frequency, and $\kappa$ is the cavity damping rate. With no loss of generality we will assume that the phase of $\epsilon$ may be chosen so that $\alpha_0$ is real. We now expand the interaction around this steady state value by making the canonical transformation $a=\bar{a}-\alpha_0$.  After the rotating wave approximation, the effective Hamiltonian in the interaction picture may then be written as
\begin{equation}
H_I=\frac{\hbar}{2}\chi(\bar{a}+\bar{a}^\dagger)b^\dagger b,
\label{ham-int}
\end{equation}
where
\begin{equation}
\chi=2G\alpha_0,
\end{equation}
and we have taken $\delta=0$ for simplicity. If we now include the damping of the cavity and the mechanical resonator, the total system state is described by the master equation
\begin{equation}
\frac{d\rho}{dt}=-\frac{i}{\hbar}[H_I,\rho]+\kappa{\cal D}[a]\rho+\gamma(\bar{N}+1){\cal D}[b]\rho+\gamma\bar{N}{\cal D}[b^\dagger]\rho,
\label{master-eqn}
\end{equation}
where ${\cal D}[c]\rho=c\rho c^\dagger -c^\dagger c\rho/2-\rho c^\dagger c/2$, $\bar{N}$ is the mean thermal occupation of the mechanical resonator bath at frequency $\nu$, $\gamma$ is the mechanical mode damping rate, and where we have dropped the bar on the $\bar{a}$ for simplicity of notation. 

The interaction Hamiltonian in equation (\ref{ham-int}) indicates that the phonon number operator for the mechanical resonator is not changed by the interaction with the optical field and, in the absence of dissipation, is a constant of the motion. The model thus describes a QND measurement of the mechanical resonator's phonon number with the cavity field forming the first stage of the measurement apparatus.  A similar model was considered in  Walls et al. \cite{WallsColMil} for both components realized as optical field modes, but only mode-a was damped.  

\section{Unconditional dynamics}

\subsection{Without mechanical damping}
As pointed out in \cite{SantamorePRB}, for direct phonon number detection the measurement apparatus must have a dynamical time scale much shorter than that of the phonon number of the relevant mechanical mode.  Therefore we begin by considering the dynamics on a time scale over which the mechanical damping can be neglected.  Such a problem is represented by the master equation above with $\gamma=0$.  The solution to this problem was given by Walls et al. in \cite{WallsColMil}:
\begin{eqnarray}
\label{WCM-soln}
\fl \rho(t) = \sum_{nm\alpha\alpha '}P_{nm}(\alpha,\alpha ')exp \left[ \frac{\chi^2}{\kappa^2}(n-m)^2(1-\kappa t/2-e^{-\kappa t/2}) \right] \nonumber \\
\nonumber \times exp \left[ (n-m)\left(-i\frac{\chi}{\kappa}\alpha '^*-i\frac{\chi}{\kappa}\alpha\right)(1-e^{-\kappa t/2}) \right] \left(|n\rangle\langle m| \right)_{(b)} \\
\otimes\frac{(|\alpha_{n}(t)\rangle\langle\alpha_{m}'(t)|)_{(a)}}{\langle\alpha_{m}'(t)|\alpha_{n}(t)\rangle},
\end{eqnarray}
where $P_{nm}(\alpha,\alpha ')$ is the initial probability distribution such that
\begin{equation}
\rho(0)=\sum_{nm\alpha\alpha '}P_{nm}(\alpha,\alpha ')(|n\rangle\langle m|)_{(b)}\otimes\frac{(|\alpha\rangle\langle\alpha'|)_{(a)}}{\langle\alpha'|\alpha\rangle},
\end{equation}
and $\alpha_{n}(t)$ and $\alpha_{m}'(t)$ are complex amplitudes corresponding to coherent states of the cavity with $\alpha$ and $\alpha '$ as the initial amplitudes:
\begin{eqnarray}
\alpha_{n}(t)  = -i\frac{\chi n}{\kappa}(1-e^{-\kappa t/2})+\alpha e^{-\kappa t/2},\\
\alpha_{m}'(t)  = -i\frac{\chi m}{\kappa}(1-e^{-\kappa t/2})+\alpha ' e^{-\kappa t/2}.
\end{eqnarray}

The solution in equation (\ref{WCM-soln}) illustrates some important features of quantum measurement. Firstly, for sufficiently strong coupling, the density operator rapidly becomes diagonal in the number basis for the resonator, which is the measured quantity. Secondly, the optical system is driven towards coherent states which are the {\em pointer basis states} \cite{Zurek} for the damped cavity field. The rate of diagonalization is proprotional to $\chi^2/\kappa$.  The resulting mixture is a classical correlation between Fock states of the resonator and corresponding pointer basis states for the field, and we find in the steady-state $\langle a \rangle(t) = -i\frac{\chi}{\kappa}\bar{n}_{b}$, where $\bar{n}_b=\langle b^\dagger b\rangle$.

\subsection{With mechanical damping}

Unravelling the full master equation with nonzero $\gamma$ is intractable, but since we are required to be in the adiabatic limit we can build our intuition of the full dynamics from the previous section.  The nature of the cavity-environment and cavity-mechanics interaction is the same for both zero and nonzero $\gamma$.  We therefore expect that the classical correlation between mechanical Fock states and cavity coherent states is preserved.  It is only the initial distribution of the mechanical Fock states that will evolve and the cavity state will adiabatically follow so that $\langle a \rangle (t) \approx -i\frac{\chi}{\kappa}\bar{n}_{b}(t)$.
We can verify this by solving the moment equations that follow directly from the master equation (\ref{master-eqn}) (which correspond to moments of the {\em unconditional state}). The moment equations are
\begin{eqnarray}
\frac{d\langle a\rangle}{dt}  =  -i\frac{\chi}{2}\bar{n}_b -\frac{\kappa}{2}\langle a\rangle,\\
\frac{d\bar{n}_b}{dt}  =  -\gamma \bar{n}_b+\gamma\bar{N}.
\end{eqnarray}
The solutions are
\begin{eqnarray}
\bar{n}_b(t)  =  \bar{n}_b(0) e^{-\gamma t}+\bar{N}(1-e^{-\gamma t}),
\label{meanphonon} \\
\fl \langle a \rangle (t)   =   \langle a\rangle(0)e^{-\kappa t/2}-i\frac{\chi}{2}\left [(\bar{n}_b(0)-\bar{N})\frac{(e^{-\gamma t}-e^{-\kappa t/2})}{\kappa/2-\gamma}+\bar{N}\frac{(1-e^{-\kappa t /2})}{\kappa/2}\right ].
\end{eqnarray}
In the adiabatic limit we recover $\langle a \rangle (t) \approx -i\frac{\chi}{\kappa}\bar{n}_{b}(t)$, as expected.  It is then apparent that we can extract information about the average phonon number by monitoring the quadrature phase amplitude to the steady state amplitude $\alpha_0$ (which has been chosen as real).  One expects that, prior to the optomechanical coupling turning on, the mechanical resonator will be in thermal equilibrium with its environment, in which case $\bar{n}_b(0)=\bar{N}$. For long times, the change in the cavity amplitude from the steady background amplitude, $\alpha_0$,  $\Delta\alpha=-i\chi\bar{N}/\kappa$. We thus regard the ratio $\chi/\kappa$ as the {\em gain} of the measurement. 

As demonstrated by equation (\ref{meanphonon}), phonon number quantum jumps are not present in the unobserved system even in the quantum limit.  Jump-like behaviour arises only in the presence of a continuous measurement, which is the subject of the next section.

\section{Conditional dynamics}

We have shown that information about the phonon number is reflected in the phase quadrature amplitude of the cavity field.  Under continuous homodyne measurement of this quadrature, the system is governed by the following stochastic master equation (SME):
\begin{eqnarray}
\label{SME}
\nonumber \fl d\rho=-\frac{i}{\hbar}[H_I,\rho]dt + \gamma(\bar{N}+1)\mathcal{D}[b]\rho dt + \gamma\bar{N}\mathcal{D}[b^\dag]\rho dt + \kappa\mathcal{D}[a]\rho dt \\
+ \sqrt{\eta\kappa}dW\mathcal{H}[ae^{-i\frac{\pi}{2}}]\rho.
\end{eqnarray}
Here, $\mathcal{D}[c]\rho$ is as defined for equation (\ref{master-eqn}), $\mathcal{H}[c]\rho=c\rho + \rho c^\dag -$ Tr$(c\rho + \rho c^\dag)$ is the measurement superoperator, $\eta$ is the detector efficiency, and $dW$ is the Wiener increment.  The homodyne measurement signal will be a photocurrent proportional to the cavity phase quadrature amplitude with stochastic noise due to the local oscillator and intrinsic quantum noise of the cavity field:
\begin{equation}
i_h(t)=\eta\kappa\langle ae^{-i\frac{\pi}{2}}+a^{\dag}e^{i\frac{\pi}{2}}\rangle + \sqrt{\eta\kappa}\xi(t),
\label{photocurrent}
\end{equation}
where $\xi(t)=dW/dt$.  As previously mentioned, we require the cavity to adiabatically follow the mechanical number state.  In this limit, we may adiabatically eliminate the cavity field from the full SME in order to reduce computational overhead.  Solving the quantum Langevin equation for the steady state of the cavity field gives $a=-i\frac{\chi}{\kappa}b^\dag b$.  The resulting SME for the density matrix of the mechanics is
\begin{equation}
\fl d\rho_b= \gamma(\bar{N}+1)\mathcal{D}[b]\rho_b dt + \gamma\bar{N}\mathcal{D}[b^\dag]\rho_b dt + \Gamma\mathcal{D}[b^\dag b]\rho_b dt +\sqrt{\eta\Gamma}\mathcal{H}[b^\dag be^{-i\pi}]\rho_b dW,
\end{equation}
where $\Gamma \equiv \frac{\chi^2}{\kappa}$.  From this it is straightforward to further simplify to the diagonal elements of $\rho_b$:
\begin{equation} 
\label{SRE}
\fl dp_n=\gamma\bar{N}[np_{n-1}-(n+1)p_n]dt + \gamma(\bar{N}+1)[(n+1)p_{n+1}-np_n]dt - 2\sqrt{\eta\Gamma}(n-\langle n \rangle)p_n dW.
\end{equation}
In the adiabatic limit the photocurrent becomes
\begin{equation}
i_h(t)=-2\eta\chi\langle b^{\dag}b \rangle_c + \sqrt{\eta\kappa}\xi(t).
\end{equation}
From hereon we set $\eta=1$.  We note that equation (\ref{SRE}) is of the same functional form as the one dealt with in Section V. of \cite{SantamorePRB}, and we refer the reader to that reference for complementary analysis and discussion.

Under continuous measurement the evolution of the system is the result of a competition between two dynamical processes.  The correlation between cavity coherent states and mechanical phonon number established in the previous section indicates that measurement of the cavity field will tend to collapse the variance of the mechanical state in the number state basis.  The thermal bath on the other hand will have the opposite effect.  For jump-like behaviour of the phonon number to arise, the measurement rate must dominate the thermalization rate.  This amounts to the {\em adiabatic} and {\em fast-measurement} conditions, which we now define.

In the absence of measurement, the thermalization rate of a mechanical Fock state $|n\rangle$ can be shown to be $\gamma[\bar{N}(n+1)+(\bar{N}+1)n]$\footnote{This expression is an upper bound on the thermalization rate; under continuous measurement the thermalization rate is decreased due to the measurement-induced collapse.} \cite{SantamorePRB}.  To be in the adiabatic limit, we therefore require the {\em adiabatic condition}: $\kappa\gg\gamma[\bar{N}(n_{max}+1)+(\bar{N}+1)n_{max}]$, where $n_{max}$ is the largest phonon number state one wishes to resolve quantum jumps from.

The {\em fast-measurement condition} arises due to the fact that even in the adiabatic limit the measurement will extract information on the phonon number at a finite rate.  The rate of number state collapse for a system described by equation (\ref{SRE})  is of the order $\Gamma$ \cite{SantamorePRB, JacobsKnight}.  The {\em fast-measurement condition} is therefore $\Gamma\gg\gamma[\bar{N}(n_{max}+1)+(\bar{N}+1)n_{max}]$.  This condition, along with the {\em adiabatic condition}, defines the "good-measurement limit," which we explore below.  We note that this limit is valid for arbitrary values of the ratio $\chi/\kappa$; the strong-coupling regime ($\chi/\kappa\gtrsim 1$) is not required.

\subsection{Good-measurement limit}

In the good-measurement limit the cavity adiabatically follows the number state of the mechanical mode, and the measurement collapses the state of the mechanical mode in the number state basis very rapidly.  Therefore we expect that the mechanical state will be a nearly pure number state $|n\rangle\langle n|$ most of the time with stochastic excitations $(n\rightarrow n+1)$ or decays $(n\rightarrow n-1)$ due to the thermal bath.  From the analysis of section III., the corresponding cavity field will have 
\begin{equation}
\langle a\rangle=\alpha_{n}(t),
\end{equation}
with a quantum jump in the phonon number,
\begin{equation}
|n\rangle\langle n|  \longrightarrow |n\pm1\rangle\langle n\pm1|,
\end{equation}
at time $t_j$ reflected as
\begin{equation}
\alpha_{n}(t) \longrightarrow \alpha_{n\pm1}(t)=-i\frac{\chi (n\pm1)}{\kappa}(1-e^{-\kappa (t-t_{j})/2})+\alpha_{n}(t_{j})e^{-\kappa (t-t_{j})/2}.
\end{equation}
The adiabatic condition ensures that the cavity amplitude will reach its steady-state value, $-i\frac{\chi}{\kappa}n$, between consecutive jump times $t_j$, and the phase quadrature amplitude will therefore trace out a step-like trajectory.

\begin{figure} [h!t]
\begin{center}
\includegraphics[scale=0.8]{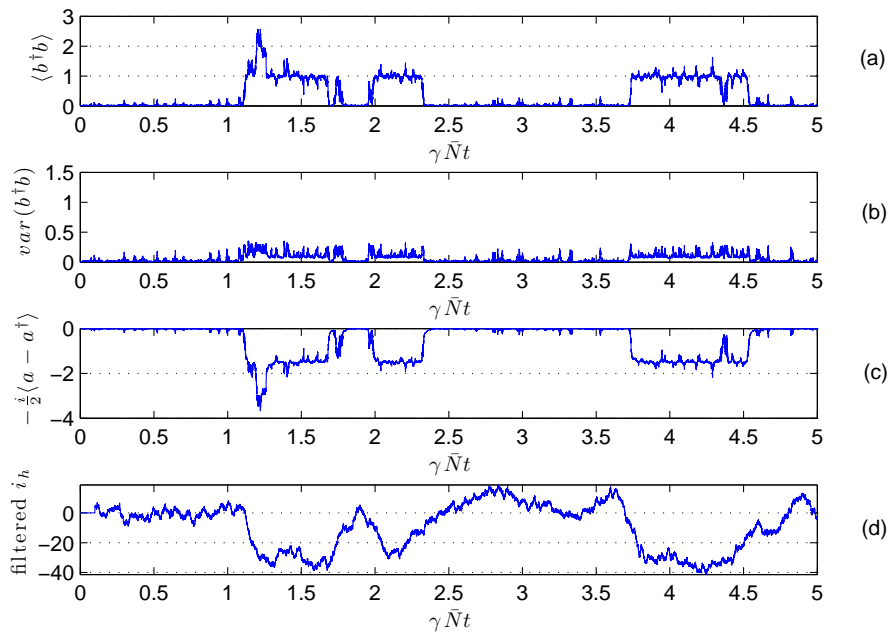}
\end{center}
\caption{Evolution of: (a) phonon number, (b) phonon number variance, and (c) cavity field phase quadrature in the good-measurement limit for low phonon numbers ($n\sim1$).  Quantum jumps in the phonon number are replicated by the cavity phase quadrature amplitude.  (d) The filtered homodyne current gives a noisy version of the cavity phase quadrature trajectory.}
\label{data28}
\end{figure}

To verify this intuition, we numerically integrate equation (\ref{SME}) with $\bar{N}=0.5$, $\chi/\kappa=1.5$, $\kappa=100\gamma\bar{N}$, and $\chi^2/\kappa=225\gamma\bar{N}$ so that we are in the good-measurement limit for low phonon numbers.  We start with the mechanics in the ground state.  The result, figure \ref{data28}, shows well-resolved quantum jumps in the phonon number replicated by the cavity phase quadrature amplitude with a gain of $\chi/\kappa=1.5$.  The small variance of the phonon number distribution indicates a high purity of the mechanical state.  The quantum trajectory formalism that we use assumes the ideal limit of an infinite amplitude local oscillator, thereby giving the photocurrent white noise with infinite amplitude in the limit that $dt \rightarrow 0$.  Numerically, the finite time step limits the noise amplitude and bandwidth.  This is qualitatively consistent with the experimental situation of a finite amplitude local oscillator and a photodetector with a finite bandwidth.  Still, the raw measurment signal, equation \ref{photocurrent}, is likely to be dominated by noise from the local oscillator.  However, we show below in section 4.4 that a simple sliding time-average of the raw signal is sufficient to reveal the quantum jumps.  We plot such a low-pass filtered version of the homodyne photocurrent in figure \ref{data28}(d). 

\subsection{Verification of adiabatic condition}

\begin{figure} [htp]
\begin{center}
\includegraphics[scale=0.8]{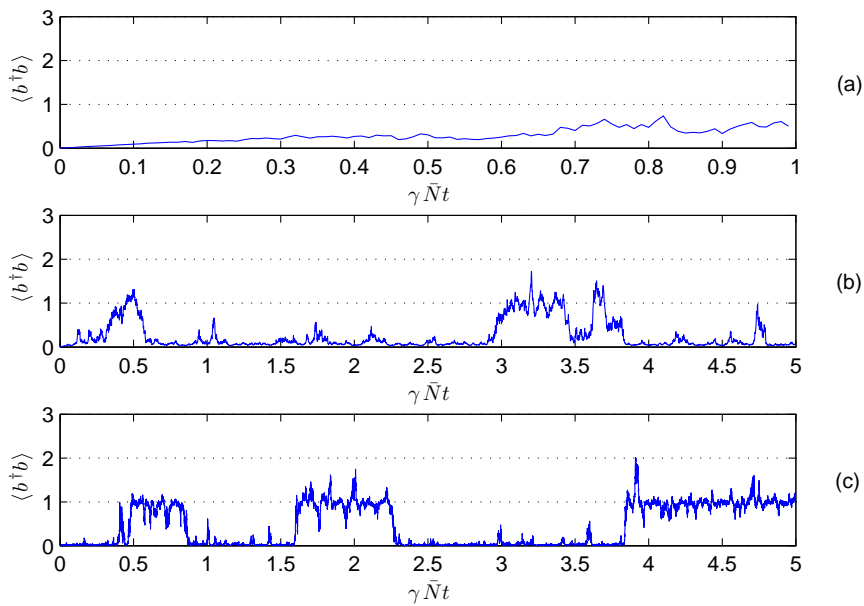}
\end{center}
\caption{Verification of adiabatic condition.  The trajectories show the evolution of the average phonon number with parameters $\bar{N}=0.5$, $\chi^2/\kappa=10^2\gamma\bar{N}$, and: (a) $\kappa=\gamma\bar{N}$, (b) $\kappa=10\gamma\bar{N}$, (c) $\kappa=10^2\gamma\bar{N}$.  Jump-like behaviour arises only when $\kappa\gg\gamma[\bar{N}(n+1)+(\bar{N}+1)n]$, where $n$ is the phonon number.}
\label{data30-32}
\end{figure}

The SME of equation (\ref{SME}) is valid outside of the adiabatic limit and we can use it to verify the necessity of the adiabatic condition ($\kappa\gg\gamma[\bar{N}(n_{max}+1)+(\bar{N}+1)n_{max}]$).  We set $\chi^2/\kappa=10^2\gamma\bar{N}$ and $\bar{N}=0.5$ so that the fast-measurement condition is satisfied for low phonon numbers, and perform simulations with $\kappa=\gamma\bar{N},10\gamma\bar{N},$ and $10^2\gamma\bar{N}$.  The mechanical mode is initialized in the ground state.  Trajectories of $\langle b^{\dag} b \rangle$ for these simulations are shown in figure \ref{data30-32}.  We see that quantum jumps  only arise when the adiabatic condition is satisfied, and thereafter are increasingly well-resolved for larger values of $\kappa$.  This simulation of figure \ref{data30-32}(a) has a shorter run time than figures \ref{data30-32}(b) and (c) due to computational limitations, but the expected steady-state is reached:  the measurement is too weak to induce number state collapse, and the mechanical mode thermalizes so that $\bar{n}_b=\bar{N}$.

\subsection{Verification of fast-measurement condition}

\begin{figure} [htp]
\begin{center}
\includegraphics[scale=0.8]{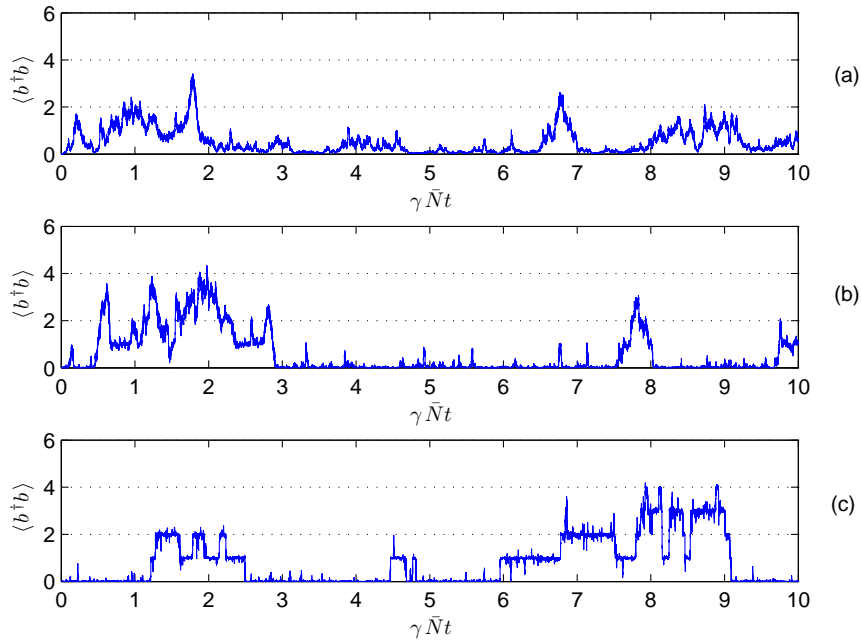}
\end{center}
\caption{Verification of fast-measurement condition.  The trajectories show the evolution of the average phonon number with parameters $\kappa=10^4\gamma\bar{N}$ and: (a) $\chi^2/\kappa=\gamma\bar{N}$, (b) $\chi^2/\kappa=10\gamma\bar{N}$, (c) $\chi^2/\kappa=10^2\gamma\bar{N}$.  Jump-like behaviour occurs only when $\chi^2/\kappa\gg\gamma[\bar{N}(n+1)+(\bar{N}+1)n]$, where $n$ is the phonon number.}
\label{fig_SRE_18-20}
\end{figure}

To verify the fast-measurement condition ($\chi^2/\kappa\gg\gamma[\bar{N}(n_{max}+1)+(\bar{N}+1)n_{max}]$) we numerically integrate equation (\ref{SRE}) with $\kappa=10^4\gamma\bar{N}$ for three cases: $\chi^2/\kappa=\gamma\bar{N}$, $\chi^2/\kappa=10\gamma\bar{N}$, and $\chi^2/\kappa=10^2\gamma\bar{N}$.  We start with the mechanics in the ground state, and the bath temperature is set at $\bar{N}=0.5$ so that the adiabatic condition is strongly satisfied for phonon numbers close to zero.  The first case, figure \ref{fig_SRE_18-20}(a), does not satisfy the fast-measurement condition and therefore does not resolve quantum jumps in the phonon number.  The second case, figure \ref{fig_SRE_18-20}(b), is on the border of the fast-measurement regime for $n\sim1$ and shows some jump-like behaviour in the phonon number.  The third case, figure \ref{fig_SRE_18-20}(c), strongly satisfies the fast-measurement condition for low phonon numbers and shows well-resolved quantum jumps in spite of being deeply within the weak coupling regime with $\chi/\kappa=10^{-1}$.

\subsection{Processing the measurement record}
We must consider that although the adiabatic and fast-measurement conditions are sufficient for phonon number quantum jumps to arise in the phonon number trajectory, that trajectory must be inferred through the measurement record, which contains additional noise.  Integration of the measurement record over a finite time interval can reduce the noise, but the time interval must be much less than the typical lifetime, $\tau_{n_{max}}\sim(\gamma[\bar{N}(n_{max}+1)+(\bar{N}+1)n_{max}])^{-1}$, of Fock state $|n_{max}\rangle$ in order for the jump-like behaviour between Fock states $\{|n\rangle : n \leq n_{max}\}$ to be resolved.  Here we show that simple averaging of the measurement record for a finite time interval $\delta t \ll \tau_{max}$ always allows one to infer the phonon number trajectory with resolution sufficient to see the quantum jumps.

We define a random variable $x$ which is the integral of the homodyne current over this interval of time:
\begin{equation}
x=i_h(t)\delta t=-2\chi \langle b^\dagger b\rangle_c\delta t+\sqrt{\kappa}dW(t).
\label{readout}
\end{equation}
This is equivalent to a generalized measurement of the number described by the conditional probability distribution
\begin{equation}
P(x|n)=\left (2\pi\kappa \delta t\right )^{-1/2}\exp\left [-\frac{(x+2\chi n \delta t )^2}{2\kappa \delta t}\right ].
\end{equation}
The probablity distribution to get the result $x$ is given by
\begin{equation}
P(x)=\sum_{n=0}^\infty p_n(t) P(x|n),
\end{equation}
where $p_n(t)$ is the phonon number distribution at the start of the time interval $\delta t$.  One can then easily verify that
\begin{eqnarray}
\bar{x}  = -2\chi\langle n\rangle \delta t,\\
var(x)  = 4\chi^2\delta t^2 \ var(n)+\kappa \delta t,
\end{eqnarray}
so that the added noise is diffusive as expected from equation (\ref{readout}).  The {\em a posteriori} state of the mechanical resonator given a particular result $x$ is given by Bayes' rule as
\begin{equation}
p_n^{(x)}(t+\delta t)=\frac{p_n(t)P(n|x)}{P(x)},
\end{equation}
where 
\begin{equation}
P(n|x)=\left (2\pi\Delta \right )^{-1/2}\exp \left [-\frac{(n-\bar{n})^2}{2\Delta}\right ],
\end{equation}
where 
\begin{eqnarray}
\bar{n}  = -\frac{x}{2\chi \delta t},\\
\Delta  = \frac{\kappa}{8\chi^2\delta t}.
\end{eqnarray}
For the conditional state to be sharply peaked on a particular number $n$ we require $\Delta \ll 1$, that is
\begin{equation}
8\Gamma \delta t \gg1,
\end{equation}
where $\Gamma = \frac{\chi^2}{\kappa}$ as above.  If we now set $\delta t \sim \tau_{n_{max}}/10$ we arrive at the condition
\begin{equation}
\Gamma\gg \gamma[\bar{N}(n_{max}+1)+(\bar{N}+1)n_{max}].
\end{equation}
This is the same as the fast-measurement condition arrived at previously.  Therefore, phonon number quantum jumps between Fock states $\{|n\rangle : n \leq n_{max}\}$ with lifetimes on the order of the typical lifetime $\tau_n$ or greater can in principle always be resolved from the measurement record by integrating the homodyne photocurrent.  This was illustrated above in figure \ref{data28}.

\subsection{Conditional phonon number statistics}

\begin{figure}[h!]
\begin{center}
\includegraphics[scale=1]{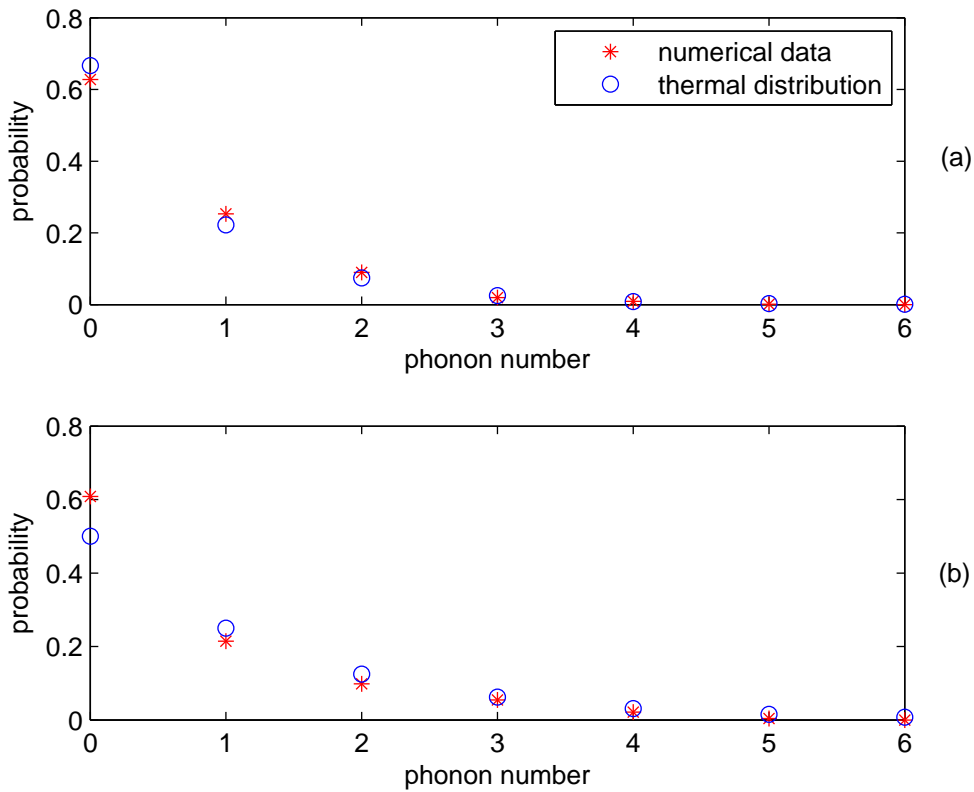}
\end{center}
\caption{Conditional statistics of the Fock state distribution when the system is well within the good-measurement limit.  (a) $\bar{N}=0.5$,  (b) $\bar{N}=1$.  }
\label{histograms}
\end{figure}

It is also interesting to look at the conditional statistics of the Fock state distribution.  We integrate equation (\ref{SRE}) with $\kappa=100\gamma\bar{N}$ and $\chi^2/\kappa=400\gamma\bar{N}$ for a total time of $30/(\gamma\bar{N})$ seconds, and bin the mean phonon number at each time step of the simulation to the nearest integer to calculate the distribution.  Figure \ref{histograms} shows the results for bath temperatures of $\bar{N}=0.5$ and $\bar{N}=1$.  The deviations in the numerical data from thermal statistics may be due to the measurement bath preventing the mechanical mode from completely thermalising with the phonon bath.

\subsection{Experimental prospects}

We discuss here the experimental prospects for observing phonon number quantum jumps in relation to the membrane-in-the-middle setup employed in \cite{Sankey}.  Assuming ground state cooling of the mechanical mode and a bath temperature much higher than the energy level spacing of the mode, the adiabatic and fast-measurement conditions respectively become $\kappa\gg\frac{k_BT}{Q\hbar}$ and $\chi^2/\kappa\gg\frac{k_BT}{Q\hbar}$, where $Q$ is the quality factor of the mechanical mode and $T$ is the bath temperature.  At $T=300mK$ the mechanical mode has $Q=1.2\times10^7$.  Thus we find the requirements for observation of the jumps to be $\kappa\gg3\times10^3 s^{-1}$ and $\chi^2/\kappa\gg3\times10^3 s^{-1}$.  The first condition is satisfied by the setup of \cite{Sankey}, which has $\kappa\sim0.3\times10^6 s^{-1}$.  Satisfying the second condition, however, remains a great challenge.  The largest coupling achieved in the setup of \cite{Sankey} is $G\sim3\times10^{-4} s^{-1}$.  With an incident power of $10 \mu W$ we find $\chi\sim10^1 s^{-1}$.  This is far short of the fast-measurement requirement $\chi^2\gg10^9 s^{-1}$, and significant further experimental effort will be required to make observation of phonon number quantum jumps feasible.  Fabricating ultra-low absorption membranes may allow for higher incident powers, thereby enhancing $\alpha_0$.  Reducing the mode frequency and the motional mass of the mode, and finding ways to further increase the $x^2$ frequency shift of the cavity can enhance $G$.  Finally, lower bath temperatures and higher $Q$ can reduce the required size of $\chi$.

\section{Conclusions}
Optomechanical systems with coupling that is quadratic in the mechanical displacement degree of freedom are promising candidates for monitoring phonon number quantum jumps in macroscopic mechanical oscillator modes.  For homodyne detection on a strongly driven single-port cavity, we have established two conditions that must be satisfied in these systems in order to observe the quantum jumps; the adiabatic condition requires the damping rate $\kappa$ of the cavity to be much larger than the thermalization rate of the mechanical mode, and the fast-measurement condition requires the quantity $\chi^2/\kappa$ to be much larger than the thermalization rate of the mechanical mode, where $\chi$ is the effective optomechanical coupling strength.  We have shown through numerical integration of the SME that satisfying these two constraints gives rise to phonon number quantum jumps in both the strong coupling ($\chi/\kappa\gtrsim1$) and weak coupling ($\chi/\kappa<1$) regimes.

\ack
GJM and TMS acknowledge support from the Australian Research Council.  AAG acknowledges support from the University of Queensland.  AAG also acknowledges discussions with J C Sankey and M J Woolley, and thanks M J Woolley for allowing adaptation of his SME integration code.

\end{document}